\documentclass[twocolumn,showkeys]{revtex4-2}
\usepackage{amsmath}
\usepackage{amssymb}
\usepackage{bm}
\usepackage{epsfig}
\usepackage{graphicx}
\usepackage{color}
\usepackage[colorlinks]{hyperref}
\usepackage[T2A]{fontenc}
\usepackage[cp1251]{inputenc}
\usepackage[english]{babel}

\begin{document}

\title{Excitons and trions with negative effective masses in two-dimensional semiconductors}

\author{M.A. Semina}
\affiliation{Ioffe Institute, 26~Polytechnicheskaya, 194021, St.-Petersburg, Russia}

\author{J.V. Mamedov}
\affiliation{National Research University, Higher School of Economics, 3A~Kantemirovskaya, 194100, St.-Petersburg, Russia}

\author{M.M. Glazov}
\affiliation{Ioffe Institute, 26~Polytechnicheskaya, 194021, St.-Petersburg, Russia}
\affiliation{National Research University, Higher School of Economics, 3A~Kantemirovskaya, 194100, St.-Petersburg, Russia}

\begin{abstract}
{We study theoretically fundamental Coulomb-correlated complexes: neutral and charged excitons, also known as trions, in transition metal dichalogenides monolayers. We focus on the situation where one of the electrons occupies excited, high-lying, conduction band characterized by a negative effective mass. We develop the theory of such high-lying excitons and trions with negative effective mass and demonstrate the key role of the non-parabolicity of the high-lying conduction band dispersion in formation of the bound exciton and trion states. We present simple, accurate and physically justified trial wavefunctions for calculating the binding energies of Coulomb-bound complexes and compare the results of variational calculations with those of a fully numerical approach. Within the developed model we discuss recent experimental results on observation of high-lying negative effective mass trions [K.-Q. Lin et al., Nat. Commun. {\bf 13}, 6980 (2022)]. }
\end{abstract}
\keywords{transition metal dichalcogenides, exciton, trion, negative effective mass, non-parabolic dispersion, mexican hat dispersion, high-lying trions}


\maketitle

\section{Introduction}\label{sec:intro}

Atomically thin transition-metal dichalcogenides (TMDC) provide a versatile platform for two-dimensional (2D) materials with tailored functionalities and fascinating physical properties~\cite{Kolobov2016book}. These semiconducting materials demonstrate outstanding optical properties -- absorption, reflection, emission -- due to excitons and trions, the Coulomb-correlated states of electrons and holes~\cite{Splendiani:2010a,Mak:2010bh,Mak:2013lh}, see Refs.~\cite{Yu30122014,Durnev_2018,RevModPhys.90.021001} for review. Controllable light-matter interaction~\cite{PhysRevLett.120.037401,Horng:19,PhysRevLett.123.067401}, ability to form van der Waals heterostructures~\cite{Geim:2013aa}{, high and tunable exciton binding energies of the excitons and trions in 2D semiconductor based systems~\cite{Semina_2022,PhysRevB.88.045318,PhysRevLett.114.107401,Courtade:2017a,PhysRevB.98.115104,Semina:2019aa}}  make these materials prime candidates for nanophotonics applications~\cite{Dufferwiel:2017aa,Schneider:2018aa,Krasnok:18}.

Usually, excitons, bound electron-hole pairs, and trions, three particle complexes formed of the electron and two holes or two electrons and a hole, involve charge carriers from the bottom conduction and topmost valence band~\cite{gross:exciton:eng,excitons:RS,ivchenko05a}. In specific cases, like bulk cuprous oxide, several excitonic series are observed that originate from closely-lying bands~\cite{0038-5670-5-2-A03,Kazimierczuk:2014yq}.  In this respect, TMDC monolayers (MLs) show unique properties. In recent experiments, the high-lying excitons and trions were observed~\cite{Lin:2021uu,Lin:2022aa} that originate from the topmost valence band holes and electrons in the excited conduction band. Corresponding optical transitions lie in the ultraviolet spectral range and can be advantageous for various applications. 

Interestingly, the effective mass of the electron in this excited conduction band is negative. It makes energy spectrum and structure of the Coulomb-correlated complexes different from that in conventional situation where the effective masses of the involved charge carriers are positive. Such situation calls for special investigation.

Here, motivated by recent experiments~\cite{Lin:2021uu,Lin:2022aa}, we study the excitons and trions where one of the charge carriers, namely, the electron, has a negative effective mass. We demonstrate the importance of  non-parabolic $k^4$ terms in the high-lying electron dispersion and present numerical and analytical results of the binding energies and wavefunctions of excitons and trions with negative-mass electrons.

The paper is organized as follows: After brief introduction (Sec.~\ref{sec:intro}) we formulate the model in Sec.~\ref{sec:model} and present the results for the excitons in Sec.~\ref{sec:excitons} and trions in Sec.~\ref{sec:trions}. Main results are summarized and a brief outlook is given in Sec.~\ref{sec:concl}.

\section{Model}\label{sec:model}

\begin{figure}[b]%
\centering
{\includegraphics[width=\linewidth]{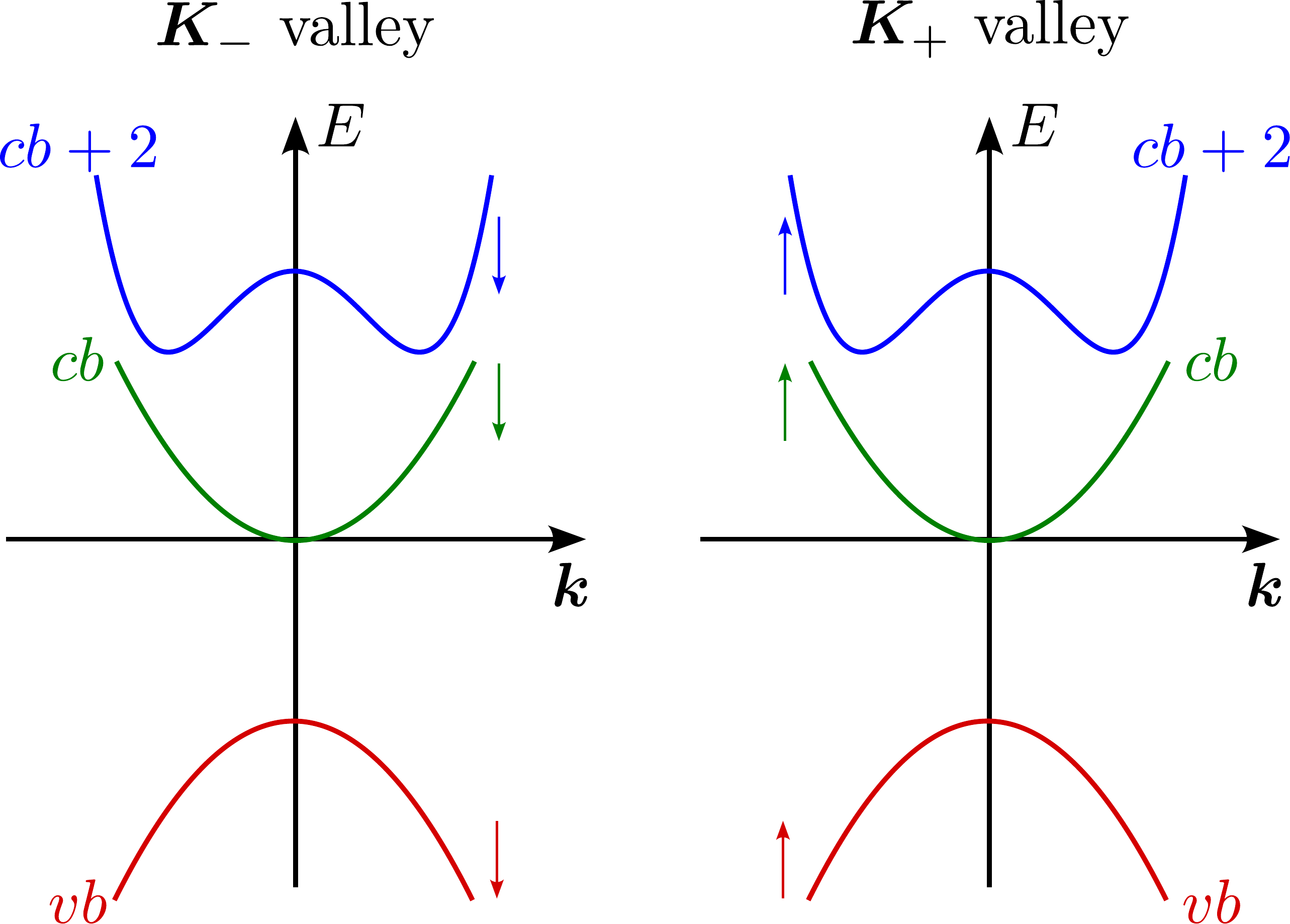}}
\caption{Schematic illustration (not to scale) of the band structure of TMDC monolayer in the vicinity of the $\bm K_\pm$ points of the Brillouin zone. The topmost valence, bottom  and high-lying conduction bands are denoted as $vb$, $cb$, and $cb+2$, respectively. Arrows denote the electron spin orientation; states with the opposite spin orientations are not shown for clarity.}\label{fig:scheme}
\end{figure}

We consider a simplified band structure of the TMDC monolayer that includes the topmost valence band $vb$, bottom conduction band $cb$ and the high-lying conduction band $cb+2$ in notations of Refs.~\cite{Durnev_2018,Lin:2021uu,Lin:2022aa,2053-1583-2-2-022001}. Figure~\ref{fig:scheme} shows schematics of the band structure  in the vicinity of the $\bm K_\pm$ points of the Brillouin zone where the direct band gap of TMDC monolayers is realized. The dispersion of the bands nearest conduction and valence bands ($vb$ and $cb$) is taken in the isotropic parabolic form:
\begin{equation}
\label{vb:cb}
E^{vb}_{k} = -E_g-\frac{\hbar^2k^2}{2m_h}, \quad  E^{cb}_{k} = \frac{\hbar^2k^2}{2m_1},
\end{equation}
 while in the dispersion of the high-lying $cb+2$ we take into account also a non-parabolic contribution in the form of the $k^4$ term:
\begin{equation}
\label{cb+2}
E^{cb+2}_{k} = E_g' + \frac{\hbar^2k^2}{2m_2} + Bk^4.
\end{equation}
Here $k$ is the electron wavevector, $E_g>0$ and $E_g'>0$ are the band gaps between $cb\leftrightarrow vb$ and $cb+2 \leftrightarrow cb$, respectively, $m_1>0$ and $m_2<0$ are, respectively,  the electron effective masses in the bottom conduction band and high-lying band, and $m_h>0$ is the effective mass of the valence band hole; the electron effective mass in $vb$ $m_{vb} = - m_h<0$. The coefficient $B>0$ describes the non-parabolic contribution to the dispersion of the high-lying electron. 

 Note that in absence of $k^4$ terms the energy $E_{k}^{cb+2}$ can become lower than $E_{k}^{cb}$ making the band notations meaningless, while $Bk^4$ renders the problem well-defined. Thus, the  dispersions~\eqref{vb:cb} and \eqref{cb+2} with $B>0$ represent a minimum model that allows us to have a consistent picture of the high-lying excitons and trions. As a result of the interplay of the $k^2$ and $k^4$ the dispersion in the $cb+2$ band has a loop (ring) of exterma at $k_* = \sqrt{-{\hbar^2}/{(4m_2 B)}}$ {demonstrating a mexican hat shape}.
In real TMDC monolayers characterized by the three-fold rotational symmetry, the dispersion of the charge carriers is anisotropic in the plane and, instead of the extrema loop, three minima can be formed. We briefly discuss the effects of anisotropy in the end of the paper. Note that a non-parabolicity in the nearest $cb$ and $vb$ is related to  the interband $\bm k\cdot \bm p$-mixing~\cite{PhysRevLett.120.187401,PhysRevB.102.155305}{, see also Ref.~\cite{PhysRevB.96.035131} for a comparative study between a single and multiband approaches;} we disregard such effects for simplicity.

To describe the excitons and trions we need to introduce the Coulomb interaction. We use it in the Rytova-Keldysh form~\cite{Rytova1967,1979JETPL..29..658K}
\begin{equation}
\label{RK}
V_{ij}(\rho) = \frac{\pi q_i q_j}{2r_0 \varkappa}\left[\mathbf H_0\left(\frac{\rho}{r_0}\right) - \mathrm Y_0 \left(\frac{\rho}{r_0}\right)\right].
\end{equation}
Here $q_{i,j}$ are the charges of the corresponding carriers ($q_e = e<0$ is the electron charge, $q_h=-e>0$ is the hole charge), $\varkappa$ is the effective dielectric constant of the environment,  $\rho$ is the interparticle distance, $r_0$ is the dielectric screening radius, $\mathbf H_0(x)$ and $\mathrm Y_0(x)$ are the Struve and Neumann functions. At large distances and/or small screening radius $\rho/r_0 \gg 1$ the potential energy takes the Coulomb form $\propto 1/\rho$, while at small distances and/or large screening radius the potential is logarithmic function of the distance $\propto \ln{\rho/r_0}$. The potential energy in the form of Eq.~\eqref{RK} is adequate for describing the Coulomb interaction in atomically thin semiconductors, see Refs.~\cite{Cudazzo:2011a,PhysRevB.88.045318,Chernikov:2014a,Courtade:2017a,PhysRevB.98.125308,Semina:2019aa} for details.

\section{Excitons}\label{sec:excitons}

We start with the theory of the two-particle bound states -- high-lying excitons (HX) -- formed from the valence band hole and high-lying electron. The effective Hamiltonian describing the relative motion of the electron and hole in the HX reads 
\begin{equation}
\label{H:HX}
\mathcal H = -\frac{\hbar^2}{2\mu_2} \Delta + B \Delta^2 + V_{eh}(\rho),
\end{equation}
where $\mu_2$ is the high-lying electron and hole reduced mass,
\begin{equation}
\label{reduced}
\mu_{1} = \frac{m_{1}m_h}{m_{1} + m_h}, \quad \mu_{2} = \frac{m_{2}m_h}{m_{2} + m_h},
\end{equation}
and $\Delta$ is the Laplace operator acting on a wavefunction  $\psi(\rho)$ with the relative electron-hole coordinate $\rho$. Since the contribution $E_g+E_{g'}$ is excluded from the Hamiltonian~\eqref{H:HX} the total energy of the high-lying exciton is $E_g+E_{g'}-E_{b,\rm HX}$ where $E_{b,\rm HX}$ is the binding energy.

We recall that in the parabolic approximation, $B=0$, the HX can be bound only if $\mu_2>0$ for attractive $V_{eh}(\rho)<0$: Indeed, the inversion of the sign of the mass can be formally considered as an inversion of the interaction potential energy sign~\cite{1971JETPL..13..229G}. Hence, for $\mu_2<0$ and $V_{eh}<0$ a bound HX state is absent. By contrast for positive $\mu_2>0$, the binding energy is given by
\begin{equation}
\label{HX:bind:parab}
E_{b,\mathrm{HX}} = \frac{2\mu_2 e^4}{\varkappa^2\hbar^2} \zeta\left(\frac{r_0\mu_2 e^2}{\varkappa \hbar^2} \right), \quad \mu_2>0,
\end{equation}
where the function $0\leqslant \zeta(x)\leqslant 1$ takes into account the dielectric screenig effect: At $x\to 0$ the function $\zeta(x) \to 1$ recovering the two-dimensional hydrogen model and at $x\to \infty$ we have $\zeta(x) \sim \ln(x)/x$~\cite{Durnev_2018,1979JETPL..29..658K}.
Interestingly, for the negative reduced mass a two-electron state can be bound despite the Coulomb repulsion between them~\cite{1971JETPL..13..229G}, see also Ref.~\cite{PhysRevB.98.115137} where the electron pairing due to the spin-orbit interaction is discussed. Note that if $\mu_2>0$, but $m_2<0$ the HX translational mass $m_{\rm HX} = m_2+m_h<0$, cf. Ref.~\cite{EFROS1984883}.

The presence of non-parabolic contribution to the dispersion $B>0$ makes HX bound for any sign and value of the reduced mass $\mu_2$, and, hence, for any value of the high-lying electron effective mass $m_2$, both positive and negative. To illustrate it we consider, instead of a Coulomb potential, a shallow short-range potential  $V_0(\rho)$
\begin{equation}
\label{shallow}
V_{eh}(\rho) < V_0(\rho) <0.
\end{equation}
The presence of the bound state for $V_0$ naturally implies the bound state for a deeper (Rytova-Keldysh) potential. For a shallow short-range interaction potential we transform the Sch\"odinger equation $\mathcal H \psi = \mathcal E\psi$ to the $k$-space and approximate the potential energy as
\[
\sum_{\bm k'} V_{0;\bm k-\bm k'} \psi_{\bm k'} \approx V_{0;0} \sum_{\bm k'} \psi_{\bm k'}, 
\]
where $V_{0,\bm q} = \int {d^2 \rho}\, V_0(\bm \rho) \exp{(\mathrm i \bm q\bm \rho)}$, $\psi_{\bm k} = \int {d^2 \rho}\, \psi(\bm \rho)\exp{(\mathrm i \bm k\bm \rho)}$ are the Fourier-components of the potential energy and wavefunction, respectively, and the normalization area is set to unity; $V_{0,0} = V_{0, \bm q=0}<0$. Thus,
\begin{equation}
\label{psi:k:scatt}
\psi_k \propto \frac{1}{\mathcal E - E_k}, 
\end{equation}
and the Schr\"odinger equation reduces to an algebraic equation; the bound state energy is found from the self-consistency requirement (see Supplementary Materials for Ref.~\cite{Lin:2022aa}):
\begin{equation}
\label{bound:scatt:0}
V_{0;0} \sum_{\bm k} \frac{1}{\mathcal E - E_k}=1, \quad E_k = Ak^2/2 + Bk^4,
\end{equation}
with $A=\hbar^2/\mu_2$. In the case of $A>0$ we obtain the bound-state energy in the form 
\begin{equation}
\label{bound:scatt:pos}
\mathcal E =  - \frac{A^2}{4B}\frac{1}{1-\exp{(-A/V_{0;0})}} \approx - \frac{A^2}{4B}e^{A/V_{0;0}},
\end{equation}
where the approximate equality holds for $V_{0;0} \to 0$. The binding energy is $E_b = -\mathcal E$. In this situation we recover exponentially shallow bound state as expected for two-dimensional system with parabolic dispersion~\cite{ll3_eng}. The non-parabolicity terms play a role of the high-momentum cut-off and determine the prefactor in the exponent in Eq.~\eqref{bound:scatt:pos}.

\begin{figure}[t]%
\centering
{\includegraphics[width=\linewidth]{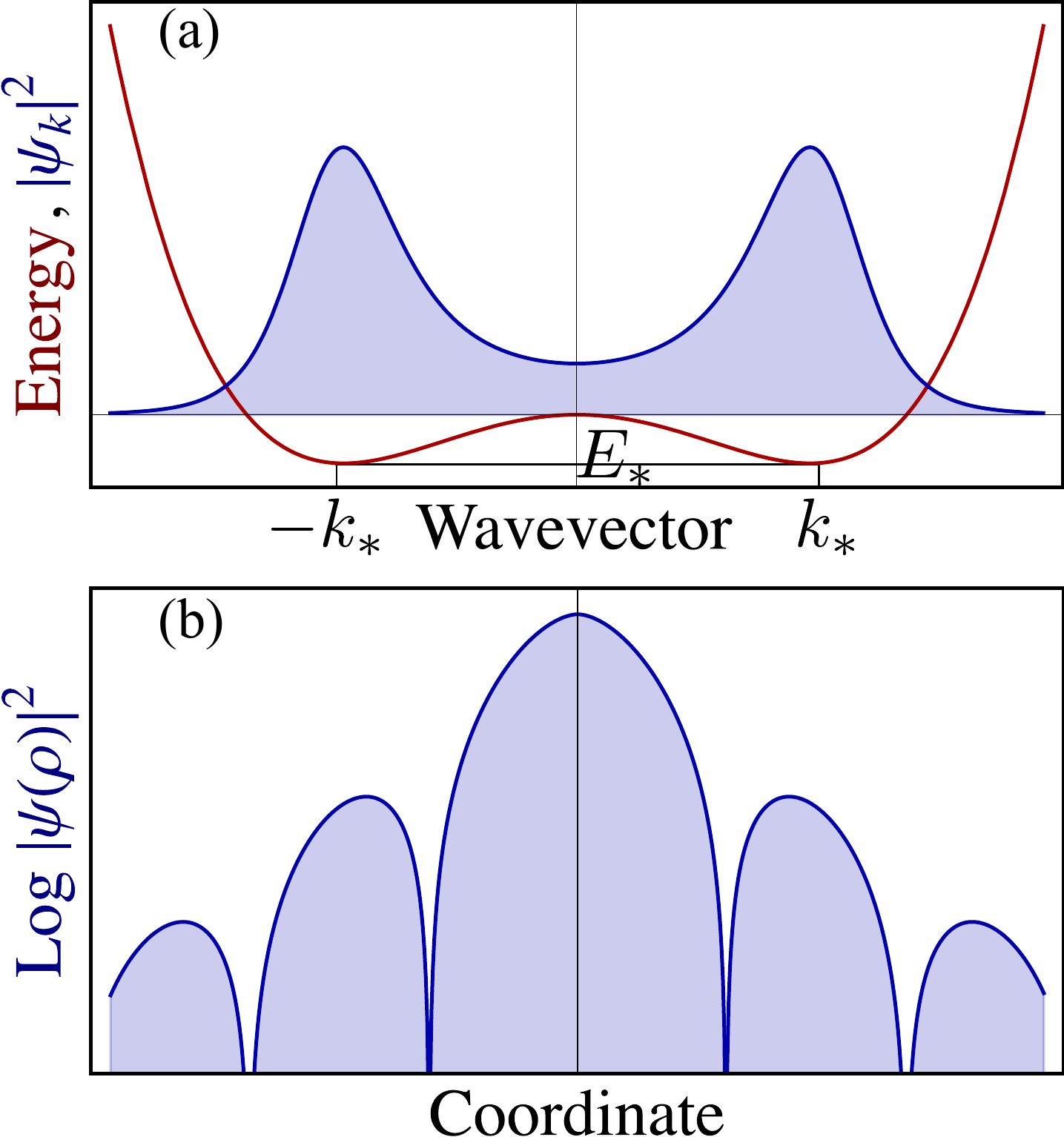}}
\caption{(a) Relative motion dispersion, Eq.~\eqref{bound:scatt:0} (dark red), and the wavefunction absolute value squared, Eq.~\eqref{psi:k:scatt} (dark blue), in the $\bm k$-space. (b) Absolute value squared of the relative motion wavefunction in the real space shown in the log-linear scale to make oscillations more pronounced. For illustrative purposes we use arb. units. {The oscillations in the real space have  the period of approximately $2\pi/k_*$.}}\label{fig:HX:wave}
\end{figure}

At $A<0$ (negative reduced mass) Eq.~\eqref{bound:scatt:0} can be transformed to the following form
\begin{equation}
\label{bound:scatt}
\arctan{\frac{A}{\sqrt{-16B \mathcal E-A^2}}} = \frac{\pi}{2}+\frac{\sqrt{-16B \mathcal E -A^2}}{2V}.
\end{equation}
The minimum of the relative motion dispersion is in this case $E_{*} = -A^2/(16B)$ corresponding to
\begin{equation}
\label{extr}
k_* = \sqrt{-\frac{A}{4 B}}.
\end{equation}
Thus the binding energy is $E_b = E_*- \mathcal E$. One can check that Eq.~\eqref{bound:scatt} has solutions with $\mathcal E<0$ for any relation between $A$ and $B$ in the reduced motion dispersion. In the important limits, 
\begin{equation}
\label{Eb:scatt}
E_{b} = 
\begin{cases}
\cfrac{(\pi V_{0,0})^2}{4B}, \quad |V_{0,0}| \ll |A|, \vspace{0.25cm}
\\
\cfrac{(\pi V_{0,0})^2}{16B}+\cfrac{AV_{0,0}}{4B}, \quad |V_{0,0}| \gg |A|.\\
\end{cases}
\end{equation}

For the negative reduced mass case the bound state is formed in the vicinity of the minima loop in the $k$-space with the relevant wavevectors $k\approx k_*$, Fig.~\ref{fig:HX:wave}(a).
Thus, as shown in Fig.~\ref{fig:HX:wave}(b), the relative motion wavefunction oscillates in the real space. Another specific feature of the wavefunctions is their behavior at $\rho \to 0$: $\psi(\rho) = {\rm const} + \rho^2 \ln{\rho}$ owing to the presence of $k^4$ terms in the dispersion. This function is sufficiently smooth at $\rho \to 0$ in contrast to the of the parabolic dispersion where the wavefunction for the shallow short-range potential well diverges as $\ln{\rho}$. 

\begin{figure}[b]%
\centering
\includegraphics[width=\linewidth]{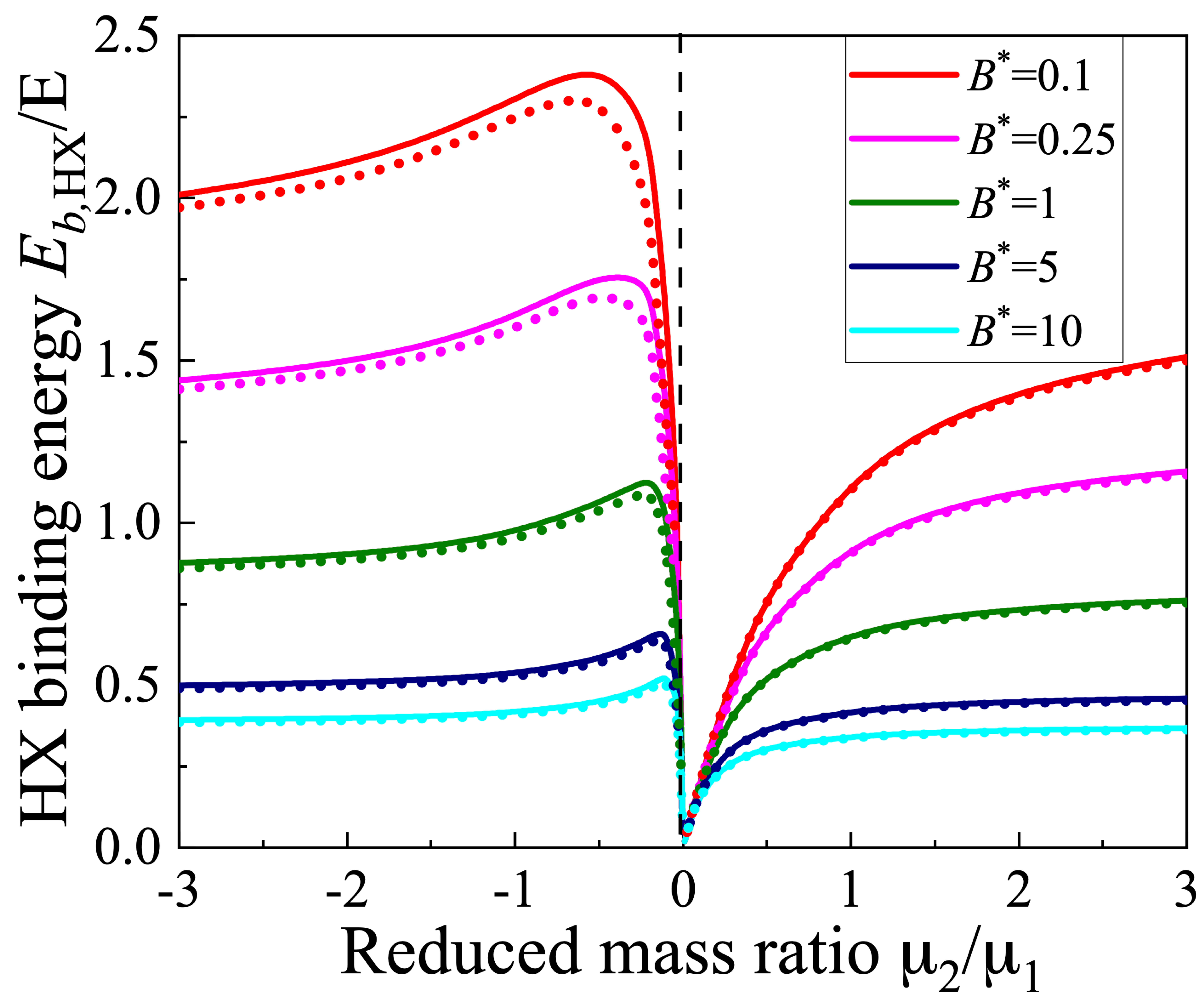}
\caption{Exciton binding energy as a function of the high-lying electron-hole reduced mass calculated for the Coulomb potential ($r_0=0$ in Eq.~\eqref{RK}) using the variational approach with the trial functions~\eqref{improved} (dots) and numerical diagonalization (solid lines). }\label{fig:HX:check}
\end{figure}

The analysis performed above forms a basis for calculating the excitonic states in the case of the Coulomb, $-e^2/(\varkappa \rho)$, and Rytova-Keldysh potential~\eqref{RK} and allows us to formulate convenient trial functions to calculate the high-lying exciton binding energy. Namely, the ground state wavefunctions both for $\mu_2>0$ and $\mu_2<0$ should behave as ${\rm const} + \rho^2$ at $\rho\to 0$, otherwise divergence occurs due to $k^4$ terms and, for $\mu_2<0$, the wavefunction should oscillate in space. Naturally, the bound state wavefunctions should decay at $\rho \to \infty$. We use the following trial functions for the HX
\begin{equation}
\label{improved}
\psi^{\pm}_{\rm HX}(\rho;a,b) \propto 
\begin{cases}
\exp{(-a\sqrt{b^2+\rho^2})}, \quad \mu_2>0,\\
J_0(a\rho) \exp{(-b\rho^2)}, \quad \mu_2<0,
\end{cases}
\end{equation}
with $a$ and $b$ being the variational parameters and superscript $\pm$ corresponds to the sign of $\mu_2$; hereafter the normalization factors are omitted.
Both functions are smooth at $\rho\to 0$, the wavefunction for $\mu_2<0$ oscillates as a function of $\rho$. We used the Bessel function $J_0(\rho)$ as it is convenient oscillating function with decaying amplitude with increase in $\rho$, which reasonably matches the oscillating behavior of the exact solution~\eqref{psi:k:scatt} in the short-range interaction model with variational parameter {$a$} controlling the period of oscillations{, see also Ref.~\cite{PhysRevB.89.041405} where detailed analytical theory of the Coulomb-bound states in two-dimensional systems with the mexican hat dispersion is presented}. We have checked accuracy of these trial functions by comparing the exciton energy found by minimizing the expectation value of the Hamiltonian~\eqref{H:HX} with the results of numerical diagonalization of Hamiltonian matrix using the non-orthogonal basis of Gaussian functions $\phi_i(\rho)=\exp(-\alpha_i\rho^2)$. Here the parameters $\alpha_i$ were taken as geometric progression. The total number $N$ of basic functions and specific values of $\alpha_i$ were chosen to optimize both the numerical convergence and computational costs~{\cite{baldereschi73,PhysRevLett.115.027402,PhysRevB.98.235401}}, typically, $N \approx 50-100$  was sufficient for excitons, further increase of $N$ did not affect the result. Note, that with the chosen basis we can obtain only exciton  ground state and axially-symmetric ($s$-shell)  excited states. In Fig.~\ref{fig:HX:check} the dotted and solid lines show $E_{b,\rm HX}$ as a function of $\mu_2$ for different values of $B$ calculated variationally (dots) and numerically. Here and in what follows we use 
\begin{equation}
\label{units} 
\mathrm E = \mu_1 e^4/(\varkappa^2 \hbar^2), \quad \mathrm a = \varkappa \hbar^2/(\mu_1 e^2),
\end{equation}
as units of the energy and length. Accordingly, the non-parabolic term in the dispersion is given by the dimensionless value $B^* =Be^4\mu_1^3/(\varkappa^2\hbar^6)$. Overall, very good agreement between the two approaches is seen. The exciton state is bound for any $\mu_2$ (positive or negative) in agreement with the analysis above.

\begin{figure}[t]%
\centering
\includegraphics[width=\linewidth]{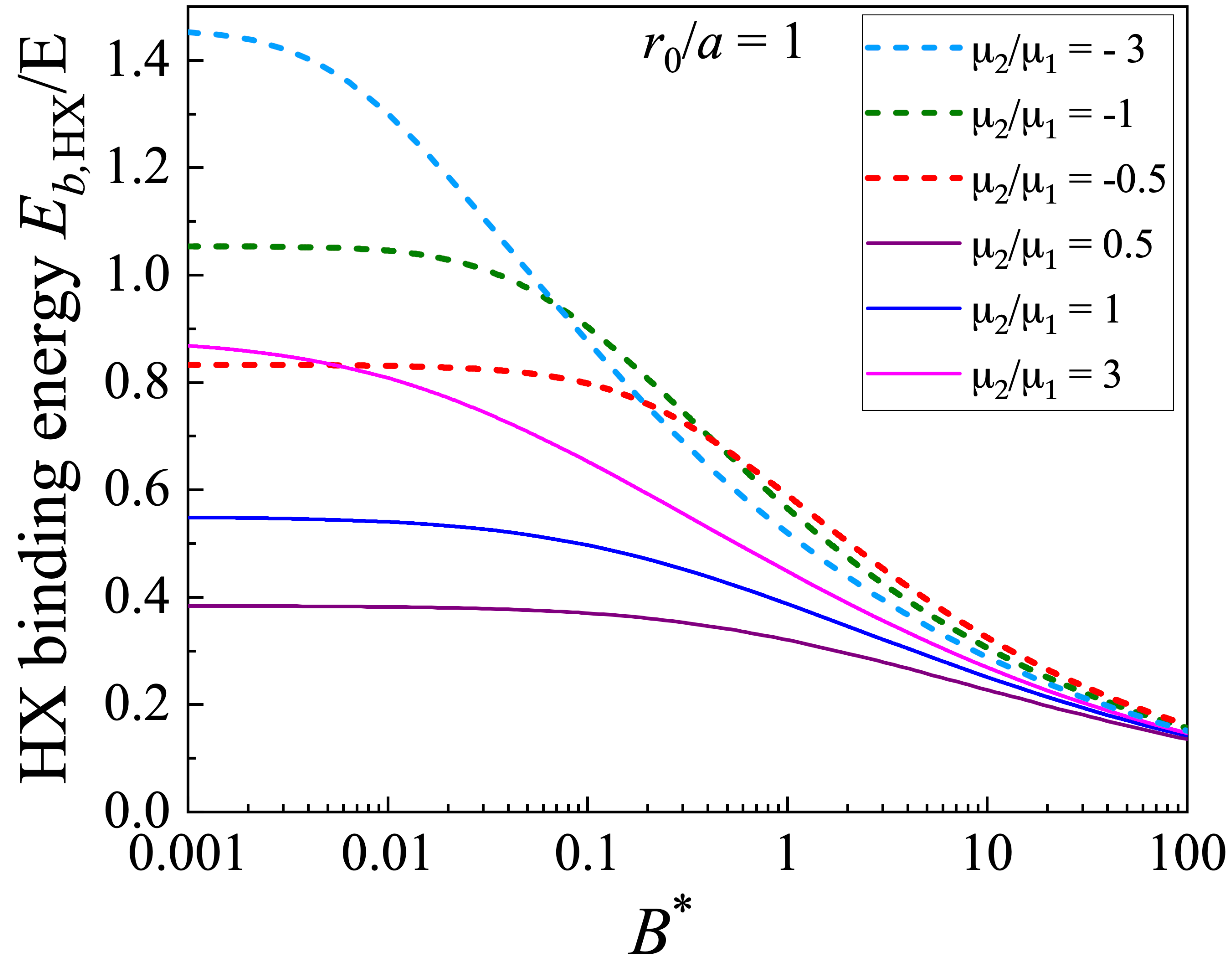}
\caption{Exciton binding energy as function of the parameter $B^*=Be^4\mu_1^3/(\varkappa^2\hbar^6)$ characterizing the non-parabolicity of the dispersion.}\label{fig:HX:res}
\end{figure}

\begin{figure}[t]%
\centering
\includegraphics[width=\linewidth]{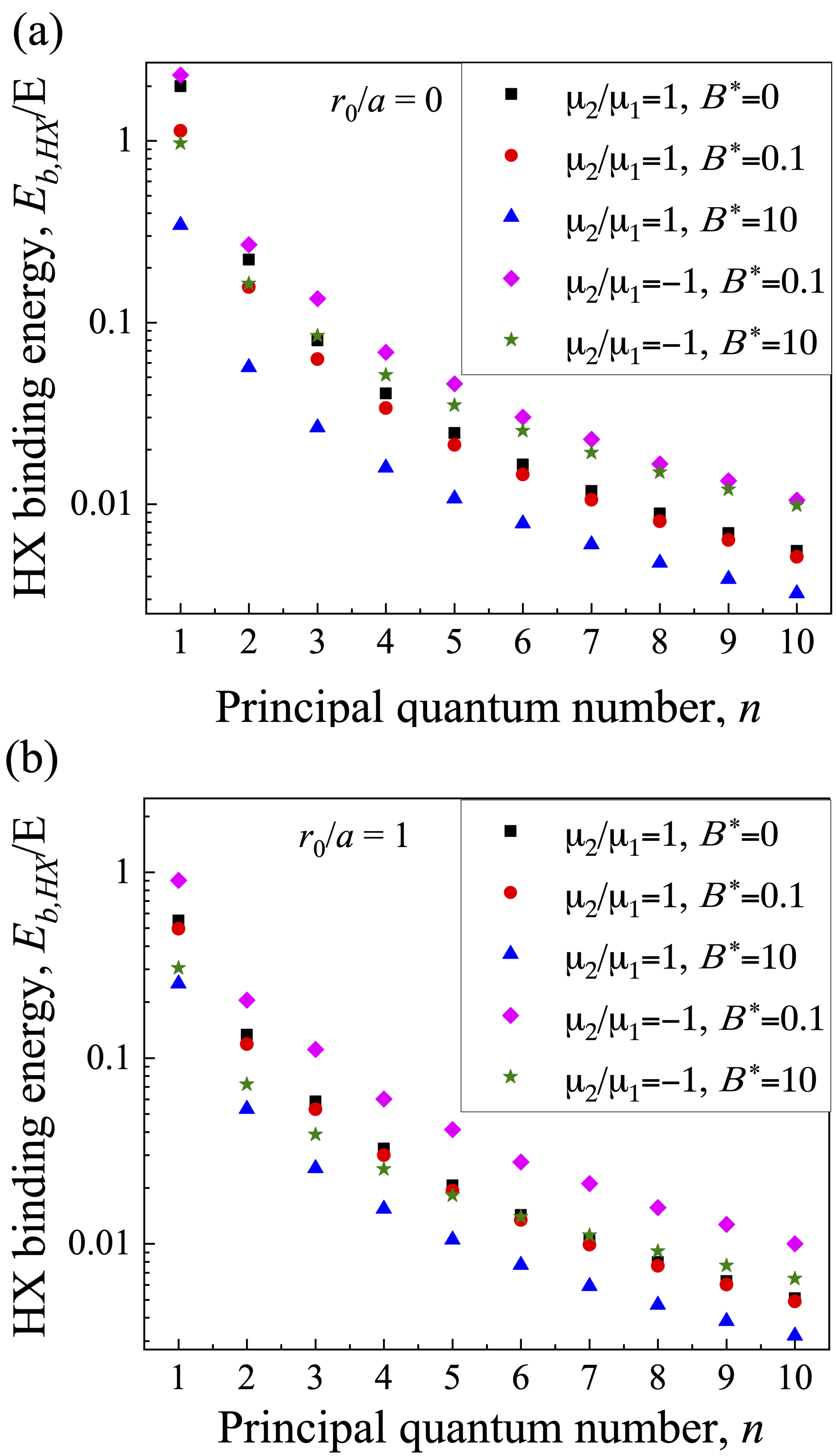}
\caption{Excitonic series for Coulomb potential ($r_0=0$ in Eq.~\eqref{RK}), panel (a), and for the screened potential ($r_0/a=1$), panel (b). $s$-shell exciton binding energies as a function of the principal quantum number $n$ are shown. }\label{fig:HX:series}
\end{figure}

Figure~\ref{fig:HX:res} shows the HX binding energy as a function the non-parabolicity parameter $B$ for several values of $\mu_2$: solid lines correspond to $\mu_2>0$ and dashed lines to $\mu_2<0$. For large $B^*$ the HX binding energy approaches the asymptotic behavior
\begin{equation}
\label{large:B}
E_{b,\rm HX} = \mathcal C \frac{\mathrm E}{(B^*)^{1/3}},
\end{equation}
with the numerical coefficient $C \approx 0.8$. The $B^{-1/3}$ power law dependence follows from the dimensional arguments taking into account that for a bound state the mean values of kinetic and potential energies of the exciton should be of the same order of magnitude and the coefficient $C$ has been found by variational approach with the Gaussian trial function. At small $B$ the HX binding energy saturates: for $\mu_2>0$ it reaches the value for the parabolic dispersion, Eq.~\eqref{HX:bind:parab} with a correction in the form {$\sim  (\mu_2/\mu_1) \mathrm E  B^* \ln{B^*}$}. The $\ln{B^*}$ factor arises because, strictly speaking, the first-order perturbation theory contribution related to the quantum mechanical average of $Bk^4$ term logarithmically diverges for hydrogenic wavefunction. Interestingly, for $\mu_2<0$ the $E_{b,\rm HX}$ also approaches a constant value that depends on $\mu_2$.  {In this case, for sufficiently small $B$ the radial motion takes place in the vicinity of the minimum in the dispersion with $k\approx k_*$ where the dispersion is parabolic and does not depend on $B$. Furthermore, the motion is essentially one-dimensional. As a result, the radial wavefunction of the exciton takes form of the bottom line in Eq.~\eqref{improved} with $a=k_*$~\cite{PhysRevB.89.041405}, resulting in the spatial oscillations with the period $\sim 2\pi/k_*$, see Fig.~\ref{fig:HX:wave}.  Hence, $E_{b,\rm HX} = \Lambda |\mu_2/\mu_1| \mathrm E $, where $\Lambda$ is a logarithmic factor that depends on the details of dispersion and screening of the potential. It is in agreement with results for the Coulomb problem in the two-dimensional electron gas with strong spin-orbit coupling~\cite{PhysRevLett.96.126402} and in the bilayer graphene~\cite{PhysRevB.89.041405} where similar dispersion can be realized~\cite{mccann:161403}}.

Finally, Fig.~\ref{fig:HX:series} shows the binding energies of HX ground and excited states for two values of $\mu_2/\mu_1=\pm 1$ and three values of $B^*$. The figure shows the energies of axially-symmetric ($s$-shell) HX states with the principal quantum numbers up to $n=10$. The effect of non-parabolicity is clearly seen. Deviations from the 2D hydrogenic model in the case of the Coulomb potential [black squares in Fig.~\ref{fig:HX:series}(a)] are clearly visible. Particularly, for positive $\mu_2$ and $B\ne 0$ the binding energies of excitonic states are smaller than for the parabolic dispersion: It is because the dispersion is steeper and hence the kinetic energy contribution which reduces the binding energy is larger. For negative $\mu_2$ the exciton energies are higher than for the parabolic case {with positive $\mu_2$}, this is because the dispersion for small $k \lesssim k_*$ is smoother.

\section{Trions}\label{sec:trions}

Now we study the high-lying trions, the three particle complexes consisting of two holes occupying the topmost valence bands and one electron in the high-lying $cb+2$ band (HX$^+$ trion) or a hole in $vb$ and two electrons one of those occupying the conduction band $cb$ and another one occupying the high-lying band $cb+2$ (HX$^-$ trion). We consider here only symmetric trions where the envelope function is symmetric with respect to the permutations of identical particles while the correspondig two-particle Bloch function is antisymmetric with respect to the permutations~\cite{Courtade:2017a}; these states are optically active at low carrier densities. Note that antisymmetric trions can also manifest themselves in the optical response but their oscillator strength is proportional to the second power of the free carrier density~\cite{PhysRevB.105.125404}. Similarly to the band edge trions where are two HX$^-$ states: intravalley (or so-called singlet) and intervalley (or triplet) ones where two electrons are, respectively, in the same valley, or in the different valleys~\cite{Courtade:2017a,Li:2020aa,He:2020aa,Robert:2021wc} resulting in the fine structure of the HX$^-$. Since the fine structure of high-lying negative trion is related to the short-range part of the electron-electron interaction [cf. Ref.~\cite{Courtade:2017a}] and, consequently, the splitting between the intra- and intervalley states is by far smaller than the trion binding energy (note that this splitting has not been observed in Ref.~\cite{Lin:2022aa}), we disregard the difference between the intra- and intervalley trions in what follows.

\subsection{Parabolic dispersion}\label{sec:trions:parab}

It is instructive to start with the parabolic dispersion model neglecting $Bk^4$ terms in the $cb+2$ dispersion. Let us consider first the HX$^+$ state. The relative motion of the holes with respect to an electron is governed by the Hamiltonian
\begin{multline}
\label{Ham:HX:+}
\mathcal H_{\mathrm{HX}^+} = - \frac{\hbar^2}{2\mu_2}\left(\Delta_1+\Delta_2 + \frac{2\sigma_2}{\sigma_2+1}  \bm \nabla_1 \bm \nabla_2\right) 
\\
+ V_{hh}(\bm \rho_1 - \bm\rho_2) + V_{eh}(\bm \rho_1) + V_{eh}(\bm \rho_2),
\end{multline}
where $\bm \rho_{i}$ are the relative coordinates of two holes ($i=1,2$) with respect to the electron, $\bm \nabla_i$ and $\Delta_i$ are the gradient and Laplace operators acting on functions of $\rho_i$, $\mu_2$ is the reduced mass of the high-lying electron and a hole, Eq.~\eqref{reduced}, and $\sigma_2 = m_h/m_2$ is the hole-to-electron mass ratio, cf. Refs.~\cite{Courtade:2017a,Sergeev:2001aa,Semina_2022}. We recall that for the neutral HX to be bound $\mu_2$ should be positive in the parabolic approximation, see Eq.~\eqref{HX:bind:parab}. In this case HX$^+$ is bound as well, since Eq.~\eqref{Ham:HX:+} describes the positive-mass situation, see~\cite{Courtade:2017a} for details. Its binding energy is a fraction of the high-lying exciton binding energy 
\begin{equation}
\label{binding:HX:+}
E_{b,\mathrm{HX}^+} = \chi E_{b,\mathrm{HX}} ,
\end{equation}
where the coefficient $\chi \sim 0.1$ depends on the screening radius $r_0$ and effective masses via $\mu_2$ and $\sigma_2$.

The situation with HX$^-$ is more involved. The relative motion Hamiltonian within a parabolic approximation takes the form
\begin{multline}
\label{Ham:HX:-}
\mathcal H_{\mathrm{HX}^-}  = -\frac{\hbar^2}{2\mu_1} \Delta_1 - \frac{\hbar^2}{2\mu_2} \Delta_2 - \frac{\hbar^2}{m_h} \bm \nabla_1 \bm \nabla_2 \\
+ V_{ee}(\bm \rho_1 - \bm\rho_2) + V_{eh}(\bm \rho_1) + V_{eh}(\bm \rho_2).
\end{multline}
In this case $\bm \rho_i$ are the relative coordinates of two electrons with respect to a hole. Taking into account that the HX$^-$ envelope function is symmetric with respect to permutation of electrons
\[
\psi_{\rm HX^-}(\bm \rho_1, \bm \rho_2) = \psi_{\rm HX^-}(\bm \rho_2, \bm \rho_1),
\]
the Hamiltonian can be mapped to the symmetrized one (cf. Eq.~\eqref{Ham:HX:+} in supplement to Ref.~\cite{Lin:2022aa})
\begin{multline}
\label{Ham:HX:-:1}
\mathcal H = - \frac{\hbar^2}{2\bar\mu}\left(\Delta_1+\Delta_2 + \frac{2\bar\sigma}{\bar\sigma+1}  \bm \nabla_1 \bm \nabla_2\right)\\
 + V_{ee}(\bm \rho_1 - \bm\rho_2) + V_{eh}(\bm \rho_1) + V_{eh}(\bm \rho_2),
\end{multline}
with the renormalized values of the parameters 
\begin{equation}
\label{mu:equiv}
\frac{1}{\bar \mu} = \frac{1}{2} \left(\frac{1}{\mu_1} + \frac{1}{\mu_2}\right),
\quad \bar \sigma = \frac{\bar \mu}{m_h - \bar \mu} = \frac{2m_1m_2}{m_h(m_1+m_2)}.
\end{equation}

Similarly to the case of the HX$^+$ one can find square-integrable eigenfunction of Hamiltonian~\eqref{Ham:HX:-:1}. However, it does not automatically mean that the corresponding negative high-lying trion is bound, since its energy can be above the energy of a neutral HX energy. Formally this is because such a trion is bound with respect to the exciton with the reduced mass $\bar\mu$ [with corresponding ``effective'' binding energy $\bar E_{b,\mathrm{HX}^-} =2\chi \bar\mu e^4/(\hbar\varkappa)^2$, cf. Eq.~\eqref{binding:HX:+}] rather than HX with the reduced mass $\mu_2$. Following Suppementary Materials to Ref.~\cite{Lin:2022aa} we obtain for the HX$^-$ binding energy 
\begin{equation}
\label{binding:HX:-:parab}
E_{b,\mathrm{HX}^-} = \frac{2\mu_2 e^4}{\hbar^2 \varkappa^2}\left[\frac{\bar \mu}{\mu_2}(1+\chi) -1\right].
\end{equation}
The binding energy should be positive, thus, in addition to $\mu_2>0$, the following conditions should hold
\begin{equation}
\label{cond}
\begin{cases}
0<m_2<m_*\equiv \dfrac{m_1m_h (1+2\chi)}{m_h-2\chi m_1}, \quad \mbox{if} \quad  m_*>0, \vspace{0.15cm} \\
0<m_2 \quad \mbox{or} \quad m_2 <m_*, \quad \mbox{if} \quad m_*<0.
\end{cases}
\end{equation}
Thus, for negative $m_2$ the condition for HX$^-$ to be bound requires $|m_2|$ to be sufficiently large. This condition can be understood from the following qualitative arguments: 
to form a bound trion state the HX considered as a rigid particle should bind with the $cb$-electron. The interaction between HX and electron is typically attractive due to both the exchange  and polarization contributions~\cite{suris:correlation,2019arXiv191204873F,PhysRevB.103.075417}. Hence, corresponding reduced mass of HX and electron should be positive yielding $m_2<-m_1-m_h<0$ where we made use of the fact that the translational mass of the HX is $m_{\rm HX} = m_2+m_h<0$ (for $\mu_2>0$) and $\mu_{e-\rm HX} = m_1 m_{\rm HX}/(m_1+m_{\rm HX})>0$.

\subsection{HX$^-$ with non-parabolic dispersion}

Next we address the effects of $cb+2$ band nonparabolicity on trions. We focus here mainly on the negatively charged high-lying trion, because this situation is particularly interesting due to an interplay of the exciton and trion binding for $\mu_2<0$. We perform two types of calculations of the HX$^-$ ground state.

The first type of calculations is variational. In the variational calculation we use symmetrized combinations of HX trial functions, Eq.~\eqref{improved} in the form
\begin{multline}
\label{trion:trial}
\psi_{\rm HX^-}(\rho_1,\rho_2;a_1,a_2,b_1,b_2) \propto \psi^\alpha_{\rm HX}(\rho_1;a_1,b_1)\psi^\beta_{\rm HX}(\rho_2;a_2,b_2)\\
 + \{ 1\leftrightarrow 2\},
\end{multline}
where $a_i$, $b_i$ ($i=1,2$) are the variational parameters, and $\alpha,\beta =\pm$ determine the particular form of the high-lying exciton wavefunction in Eq.~\eqref{improved}: For $\mu_2>0$ we use $\alpha=\beta=+$, while for $\mu_2<0$ we use $\alpha=+$ and $\beta=-$. In the latter case such sign convention allows us to take into account that one of the electrons in the HX$^-$ (from $cb$) has a positive effective mass and the other one (from $cb+2$) has a negative mass such that reduced mass $\mu_2<0$. We have also checked a trial function with both $\alpha=\beta=-$ for trions with the negative reduced mass, $\mu_2<0$, and the resulting energies were very close to obtained with function with $\alpha=+$ and $\beta=-$.

\begin{figure}[b]%
\centering
\includegraphics[width=\linewidth]{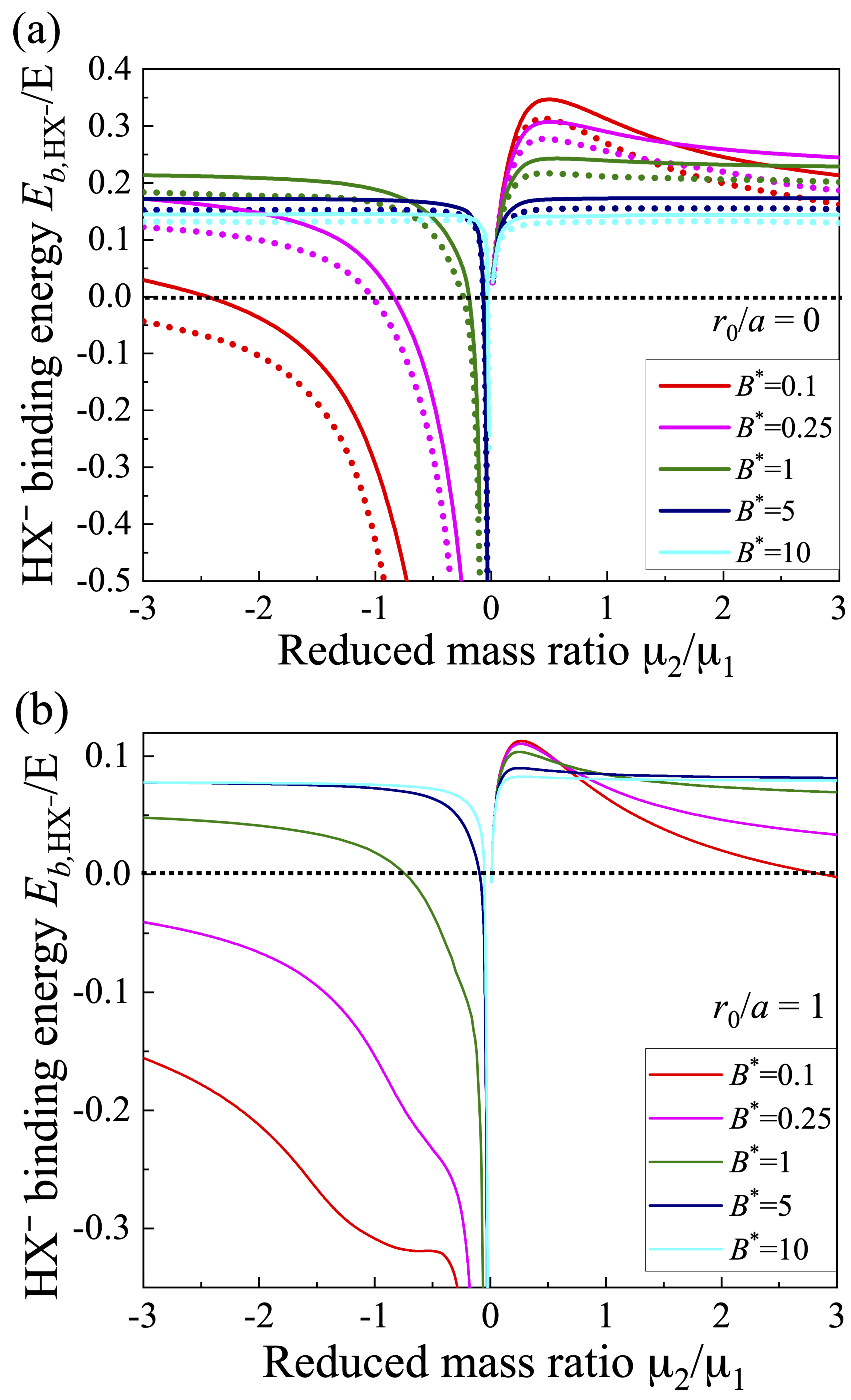}
\caption{HX$^-$ binding energy as a function of the high-lying electron-hole reduced mass calculated for the Coulomb potential (a) and for screened potential (b). Dotted lines in (a) show the results of variational calculation and solid lines [in panels (a) and (b)] show the results of full numerical approach.  }\label{fig:HX-:mu}
\end{figure}

The second type of calculations is used to test the variational approach and provide more accurate numerical framework for determining the high-lying trion states. In this calculation the HX$^-$ wavefunction is decomposed as 
\begin{multline}\label{trion:num}
\psi_{\rm HX^-}(\bm\rho_1,\bm\rho_2)\\
=\sum_{i,j,k} C_{ijk} \left(e^{-\alpha_i\rho_1^2-\beta_j\rho_2^2}+e^{-\alpha_i\rho_2^2-\beta_j\rho_1^2}\right)e^{-\delta_k|\bm\rho_1-\bm\rho_2|^2},
\end{multline}
where $\alpha_i,\beta_j,\delta_k$ are parameters whose values were taken as geometric progression. Similarly to calculation of the high-lying excitons presented above, the total number of basic functions and specific values of $\alpha_i,\beta_j,\delta_k$ were chosen for the best combination of convergence and computational costs. The coefficients $C_{ijk}$ were determined by minimizing the total energy. The wavefunction~\eqref{trion:num} provides  rather accurate form of the radial wavefunctions for relative motion of electrons and, importantly, takes into account, via the factor $\exp{(-\delta_k|\bm\rho_1-\bm\rho_2|^2)}$, correlation between the electron motion.

Figure~\ref{fig:HX-:mu} shows the dependence of the HX$^-$ binding energy on the high-lying electron to hole reduced mass $\mu_2$ calculated for several values of the non-parabolicity parameter $B$. Solid lines show the results of the full numerical calculation, while dotted lines in Fig.~\ref{fig:HX-:mu}(a) demonstrate the results of the variational approach with the trial functions~\eqref{trion:trial}. The variational calculation gives reasonable estimate of the binding energy being by $10\%\ldots 30\%$ lower than the ``exact'' value found using the wavefunction~\eqref{trion:num}. We have also performed the numerical calculation with the function in the form of Eq.~\eqref{trion:num} but without correlation factors, i.e., setting $\delta_k \equiv 0$. These results turn out to be almost indistinguishable from the results of variational calculation, which justifies the choice of the trial functions~\eqref{trion:trial} for the variational calculation. 

\begin{figure}[t]%
\centering
\includegraphics[width=\linewidth]{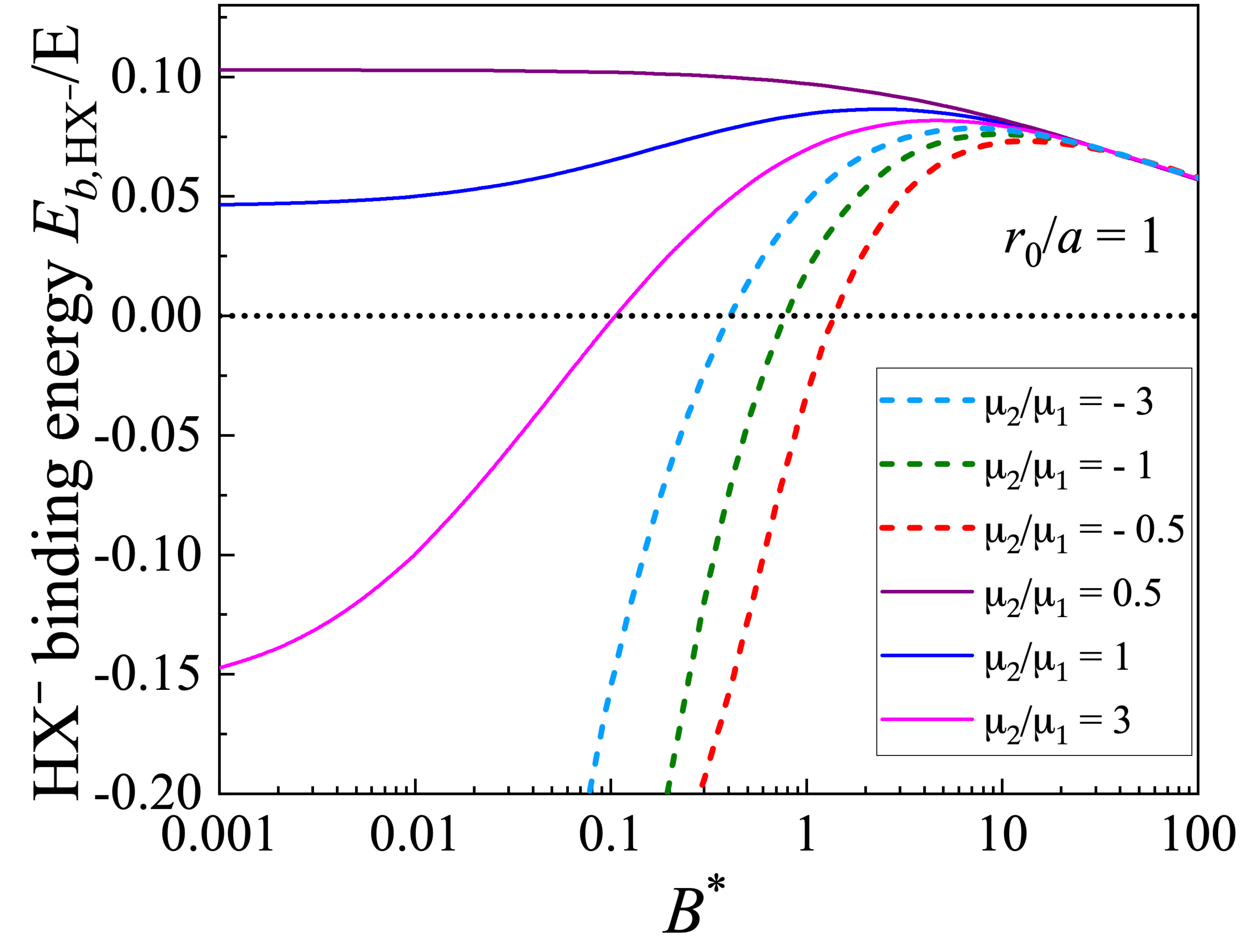}
\caption{High-lying trion binding energy as a function of non-parabolicity parameter $B^*$. }\label{fig:HX-:B}
\end{figure}

For negative $\mu_2$ the larger is $|\mu_2|$, the larger is the high-lying trion binding energy. For positive $\mu_2$ a maximum in the dependence of $E_{b,\mathrm{HX}^-}$ on $\mu_2$ is seen for small $B$. Qualitatively, this maximum can be understood within the framework of the analytical expression for the HX$^-$ binding energy in the parabolic approximation, Eq.~\eqref{binding:HX:-:parab}. It appears as a result of an interplay of two terms: the first, positive term, weakly increases with increase in $\mu_2$, while the absolute value of the second, negative term, increases linearly with $\mu_2$. This maximum becomes more pronounced in the case of the screened Rytova-Keldysh potential, Fig.~\ref{fig:HX-:mu}(b).

The dependence of the high-lying trion binding energy on the non-parabolic contribution to the dispersion characterized by the parameter $B$ is shown in Fig.~\ref{fig:HX-:B}. For small $B$ ($B^* \ll 1$) the HX$^-$ binding energy increases with increase in $B$ and strongly depends on $\mu_2$. Hence, the presence of $k^4$ terms in the high-lying electron dispersion makes trions more stable. For large non-parabolic term ($B^* \gg 1$) the $E_{b,\mathrm{HX}^-}$ decreases with increasing $B$ regardless the value of $\mu_2$ following the same $B^{-1/3}$ power law as for the HX, Eq.~\eqref{large:B} with a different coefficient yielding relatively large high-lying trion-to high-lying exciton binding energy ratio
\begin{equation}
\label{ratio:large:B}
\frac{E_{b,\rm HX^-}}{E_{b,\rm HX}} \approx {0.3}.
\end{equation}  

To summarize, the presence of $k^4$ terms in the high-lying electron dispersion significantly expand the range $\mu_2/\mu_1$ where the high-lying HX$^-$ trion is bound. As for positive trion, HX$^+$ it is bound already in the parabolic approximation and our estimates show that it remains bound in the presence of non-parabolic contributions to the dispersion.

\subsection{Discussion of the results}

Let us now briefly discuss the obtained results in view of experimental data reported in Ref.~\cite{Lin:2022aa}. For rough estimates we note that in TMDC monolayers the valence band hole and the conduction band ($cb$) electron effective masses are about the same, $m_h \approx m_1$. The electron effective mass in $cb+2$ conduction band has a similar absolute value, but it is negative. Estimates based on the DFT approach presented in Ref.~\cite{Lin:2021uu} show that for WSe$_2$ monolayers $|m_2| \approx 0.46m_0 >m_h \approx 0.36m_0$ with $m_0$ being the free-electron mass, making the reduced mass $\mu_2\approx 1.66m_0>0$ in Eq.~\eqref{reduced}. Additional evidence for $|m_2|>m_h$ follows from the strong phonon progression of HX observed in Ref.~\cite{Lin:2021uu} indicating that the translational mass of the high-lyign exciton is negative. Thus, neutral high-lying exciton, HX, is bound even if $Bk^4$ terms are neglected in the $cb+2$ dispersion. For such parameters, however, $m_*>0$ in Eq.~\eqref{cond} (for $m_1 \approx m_h$ and reasonable $\chi \approx 0.2$ the {$m_* \approx 2.3m_1$}). Hence, according to Eq.~\eqref{cond} the HX$^-$ is not bound in the parabolic approximation. Thus, we come to the conclusion that $Bk^4$ contribution should be sizeable to make HX$^-$ bound. 

Note that experimentally observed HX$^-$ binding energies are $\approx 43$~meV for WSe$_2$ monolayer and $\approx 21$~meV for MoSe$_2$ monolayer~\cite{Lin:2022aa}. In the former case it is slightly larger than the band-edge trion, X$^-$, binding energy, while in the latter case it is slightly smaller than that of X$^-$. Depending on $\mu_2/\mu_1$ and $B$ the HX$^-$ binding energy can be on the order of $10\%\ldots 25\%$ of the HX binding energy, thus, somewhat increased values of $E_{b,\rm HX^-}$ compared to $E_{b,\mathrm X^-}$ can be related to (i) enhancement of the $E_{b,\rm HX}$ due to rather large $\mu_2/\mu_1 \approx 6\ldots 10$ for the estimated parameters and (ii) to large $B^*$ where, as mentionned above, the trion-to-exciton binding energy ratio turns out to be quite large, Eq.~\eqref{ratio:large:B}. 

In this theoretical paper we abstain from further analysis of the experimental data and more detailed comparison of the calculations with experiment. The main reason is that the dispersion in $cb+2$ band is quite complicated~\cite{2053-1583-2-2-022001,Lin:2021uu} and contains, in addition to simplified Eq.~\eqref{cb+2}, anisotropic terms. Also, experimental data on high-lying excitons binding energy are not available to the best of our knowledge. Still our theoretical results in combination with experimental data~\cite{Lin:2022aa} indicate importance of the non-parabolic terms in the high-lying electron dispersion for formation of the three-particle Coulomb complexes.

\section{Conclusion and outlook}\label{sec:concl}

To conclude, we have developed the theory of  high-lying excitons and trions in two-dimensional semiconductors. Such Coulomb-bound complexes involve one electron in the excited conduction band with the negative effective mass and a non-parabolic dispersion. We have developed (i) variational method for calculating such complexes with simple and physically justified trial functions and (ii) the efficient and accurate numerical approach based on decomposition of the wavefunctions of Gaussians. We have demonstrated the importance of the band non-parabolicity for formation of the high-lying excitons and trions. In particular, for negative reduced mass the presence of $k^4$ terms in the high-lying electron dispersion makes exciton bound and strongly enhances the range of stability of the negatively charged high-lying trions. Our estimates show that the high-lying trion binding energies can be in the range of $10\%\ldots 30\%$ of the high-lying exciton binding energy, i.e., on the order of several tens of meV for transition-metal dichalcogenide monolayers.

The developed theory is not limited to the monolayer transition-metal dichalcogenides. {In the gapped bilayer graphene the dispersion contains the mexican hat features both in conduction and valence bands~\cite{mccann:161403} and tunable excitons are observed in this material~\cite{Ju:2017aa}.} In several other material platforms, including few-layer Ga- and In-monoseledines and monosulfides ~\cite{PhysRevB.90.235302} the dispersion with a ring of extrema in the valence band. Similar situation is probably realized in two-dimensional hexagonal BN~\cite{PhysRevLett.96.026402,Cassabois:2016aa}. In this respect, we can also mention bulk GaP with the camel's back dispersion~\cite{Altarelli19781101,Glinskii1979631}. Importantly, dispersion engineering, e.g., in moire lattices can be used to realize the non-parabolic dispersion with an extremum ring.  In this regard, developed theoretical approaches will be helpful for studying the fundamental quasiparticles, excitons and trions, in a wide range of material systems.




\section*{Acknowledgments}
The authors are grateful to Kai-Qiang Lin, Jonas D. Ziegler, Alexey Chernikov{, and Boris I. Shklovskii} for valuable discussions. This work was supported by the State Task of Ioffe Institute.


\bibliographystyle{unsrt}


\begin{thebibliography}{60}

\bibitem{Kolobov2016book}
Alexander~V. Kolobov and Junji Tominaga.
\newblock {\em Two-Dimensional Transition-Metal Dichalcogenides}.
\newblock Springer International Publishing, 2016.

\bibitem{Splendiani:2010a}
Andrea Splendiani, Liang Sun, Yuanbo Zhang, Tianshu Li, Jonghwan Kim, Chi-Yung
  Chim, Giulia Galli, and Feng Wang.
\newblock Emerging photoluminescence in monolayer {MoS}$_2$.
\newblock {\em Nano Letters}, 10:1271, 2010.

\bibitem{Mak:2010bh}
Kin~Fai Mak, Changgu Lee, James Hone, Jie Shan, and Tony~F. Heinz.
\newblock Atomically thin {MoS}$_{2}$: A new direct-gap semiconductor.
\newblock {\em Phys. Rev. Lett.}, 105:136805, 2010.

\bibitem{Mak:2013lh}
Kin~Fai Mak, Keliang He, Changgu Lee, Gwan~Hyoung Lee, James Hone, Tony~F.
  Heinz, and Jie Shan.
\newblock Tightly bound trions in monolayer {MoS}$_2$.
\newblock {\em Nat Mater}, 12(3):207--211, 2013.

\bibitem{Yu30122014}
Hongyi Yu, Xiaodong Cui, Xiaodong Xu, and Wang Yao.
\newblock Valley excitons in two-dimensional semiconductors.
\newblock {\em National Science Review}, 2(1):57--70, 2015.

\bibitem{Durnev_2018}
M~V Durnev and M~M Glazov.
\newblock Excitons and trions in two-dimensional semiconductors based on
  transition metal dichalcogenides.
\newblock {\em Physics-Uspekhi}, 61(9):825--845, 2018.

\bibitem{RevModPhys.90.021001}
Gang Wang, Alexey Chernikov, Mikhail~M. Glazov, Tony~F. Heinz, Xavier Marie,
  Thierry Amand, and Bernhard Urbaszek.
\newblock Colloquium: Excitons in atomically thin transition metal
  dichalcogenides.
\newblock {\em Rev. Mod. Phys.}, 90:021001, 2018.

\bibitem{PhysRevLett.120.037401}
Patrick Back, Sina Zeytinoglu, Aroosa Ijaz, Martin Kroner, and Atac
  Imamo\ifmmode~\breve{g}\else \u{g}\fi{}lu.
\newblock {Realization of an Electrically Tunable Narrow-Bandwidth Atomically
  Thin Mirror Using Monolayer MoSe$_{2}$}.
\newblock {\em Phys. Rev. Lett.}, 120:037401, 2018.

\bibitem{Horng:19}
Jason Horng, Yu-Hsun Chou, Tsu-Chi Chang, Chu-Yuan Hsu, Tien-Chang Lu, and Hui
  Deng.
\newblock Engineering radiative coupling of excitons in 2d semiconductors.
\newblock {\em Optica}, 6(11):1443--1448, 2019.

\bibitem{PhysRevLett.123.067401}
H.~H. Fang, B.~Han, C.~Robert, M.~A. Semina, D.~Lagarde, E.~Courtade,
  T.~Taniguchi, K.~Watanabe, T.~Amand, B.~Urbaszek, M.~M. Glazov, and X.~Marie.
\newblock {Control of the Exciton Radiative Lifetime in van der Waals
  Heterostructures}.
\newblock {\em Phys. Rev. Lett.}, 123:067401, 2019.

\bibitem{Geim:2013aa}
A.~K. Geim and I.~V. Grigorieva.
\newblock Van der $\mbox{W}$aals heterostructures.
\newblock {\em Nature}, 499(7459):419--425, 2013.

\bibitem{Semina_2022}
M~A Semina and R~A Suris.
\newblock Localized excitons and trions in semiconductor nanosystems.
\newblock {\em Physics-Uspekhi}, 65(2):111--130, 2022.

\bibitem{PhysRevB.88.045318}
Timothy~C. Berkelbach, Mark~S. Hybertsen, and David~R. Reichman.
\newblock Theory of neutral and charged excitons in monolayer transition metal
  dichalcogenides.
\newblock {\em Phys. Rev. B}, 88:045318, 2013.

\bibitem{PhysRevLett.114.107401}
{Bogdan Ganchev, Neil Drummond, Igor Aleiner, and Vladimir Fal'ko.
\newblock Three-particle complexes in two-dimensional semiconductors.
\newblock {\em Phys. Rev. Lett.}, 114:107401, 2015.}

\bibitem{Courtade:2017a}
E.~Courtade, M.~Semina, M.~Manca, M.~M. Glazov, C.~Robert, F.~Cadiz, G.~Wang,
  T.~Taniguchi, K.~Watanabe, M.~Pierre, W.~Escoffier, E.~L. Ivchenko,
  P.~Renucci, X.~Marie, T.~Amand, and B.~Urbaszek.
\newblock Charged excitons in monolayer {WSe}$_{2}$: Experiment and theory.
\newblock {\em Phys. Rev. B}, 96:085302, 2017.

\bibitem{PhysRevB.98.115104}
{M.~Van~der Donck and F.~M. Peeters.
\newblock Interlayer excitons in transition metal dichalcogenide
  heterostructures.
\newblock {\em Phys. Rev. B}, 98:115104, 2018.}

\bibitem{Semina:2019aa}
M.~A. Semina.
\newblock Excitons and trions in bilayer van der waals heterostructures.
\newblock {\em Physics of the Solid State}, 61(11):2218--2223, 2019.

\bibitem{Dufferwiel:2017aa}
S.~Dufferwiel, T.~P. Lyons, D.~D. Solnyshkov, A.~A.~P. Trichet, F.~Withers,
  S.~Schwarz, G.~Malpuech, J.~M. Smith, K.~S. Novoselov, M.~S. Skolnick, D.~N.
  Krizhanovskii, and A.~I. Tartakovskii.
\newblock Valley-addressable polaritons in atomically thin semiconductors.
\newblock {\em Nature Photonics}, 11(8):497--501, 2017.

\bibitem{Schneider:2018aa}
Christian Schneider, Mikhail~M. Glazov, Tobias Korn, Sven H{\"o}fling, and
  Bernhard Urbaszek.
\newblock Two-dimensional semiconductors in the regime of strong light-matter
  coupling.
\newblock {\em Nature Communications}, 9(1):2695, 2018.

\bibitem{Krasnok:18}
Alex Krasnok, Sergey Lepeshov, and Andrea Al\'{u}.
\newblock Nanophotonics with 2d transition metal dichalcogenides.
\newblock {\em Opt. Express}, 26(12):15972--15994, 2018.

\bibitem{gross:exciton:eng}
E.~F. Gross and N.~A. Karrjew.
\newblock Light absorption by cuprous oxide crystal in infrared and visible
  part of the spectrum.
\newblock {\em Dokl. Akad. Nauk SSSR}, 84:471, 1952.

\bibitem{excitons:RS}
E.~I. Rashba and M.~D. Sturge, editors.
\newblock {\em Excitons}.
\newblock North-Holland Publishing Company, 1982.

\bibitem{ivchenko05a}
E.~L. Ivchenko.
\newblock {\em Optical spectroscopy of semiconductor nanostructures}.
\newblock Alpha Science, Harrow UK, 2005.

\bibitem{0038-5670-5-2-A03}
E~F Gross.
\newblock Excitons and their motion in crystal lattices.
\newblock {\em Soviet Physics Uspekhi}, 5(2):195, 1962.

\bibitem{Kazimierczuk:2014yq}
T.~Kazimierczuk, D.~Frohlich, S.~Scheel, H.~Stolz, and M.~Bayer.
\newblock {Giant Rydberg excitons in the copper oxide Cu$_2$O}.
\newblock {\em Nature}, 514(7522):343--347, 2014.

\bibitem{Lin:2021uu}
Kai-Qiang Lin, Chin~Shen Ong, Sebastian Bange, Paulo~E. Faria~Junior, Bo~Peng,
  Jonas~D. Ziegler, Jonas Zipfel, Christian B{\"a}uml, Nicola Paradiso, Kenji
  Watanabe, Takashi Taniguchi, Christoph Strunk, Bartomeu Monserrat, Jaroslav
  Fabian, Alexey Chernikov, Diana~Y. Qiu, Steven~G. Louie, and John~M. Lupton.
\newblock {Narrow-band high-lying excitons with negative-mass electrons in
  monolayer WSe$_2$}.
\newblock {\em Nature Communications}, 12(1):5500, 2021.

\bibitem{Lin:2022aa}
Kai-Qiang Lin, Jonas~D. Ziegler, Marina~A. Semina, Javid~V. Mamedov, Kenji
  Watanabe, Takashi Taniguchi, Sebastian Bange, Alexey Chernikov, Mikhail~M.
  Glazov, and John~M. Lupton.
\newblock High-lying valley-polarized trions in 2d semiconductors.
\newblock {\em Nature Communications}, 13(1):6980, 2022.

\bibitem{2053-1583-2-2-022001}
Andor Kormanyos, Guido Burkard, Martin Gmitra, Jaroslav Fabian, Viktor
  Z{\'o}lyomi, Neil~D Drummond, and Vladimir Fal'ko.
\newblock $\bm k\cdot \bm p$ theory for two-dimensional transition metal
  dichalcogenide semiconductors.
\newblock {\em 2D Materials}, 2(2):022001, 2015.

\bibitem{PhysRevLett.120.187401}
Maxim Trushin, Mark~Oliver Goerbig, and Wolfgang Belzig.
\newblock Model prediction of self-rotating excitons in two-dimensional
  transition-metal dichalcogenides.
\newblock {\em Phys. Rev. Lett.}, 120:187401, 2018.

\bibitem{PhysRevB.102.155305}
N.~V. Leppenen, L.~E. Golub, and E.~L. Ivchenko.
\newblock {Exciton oscillator strength in two-dimensional Dirac materials}.
\newblock {\em Phys. Rev. B}, 102:155305, 2020.

\bibitem{PhysRevB.96.035131}
{M.~Van~der Donck, M.~Zarenia, and F.~M. Peeters.
\newblock Excitons and trions in monolayer transition metal dichalcogenides: A
  comparative study between the multiband model and the quadratic single-band
  model.
\newblock {\em Phys. Rev. B}, 96:035131, 2017.}

\bibitem{Rytova1967}
N.~S. Rytova.
\newblock {Screened potential of a point charge in a thin film}.
\newblock {\em Proc. MSU, Phys., Astron.}, 3:18, 1967.

\bibitem{1979JETPL..29..658K}
L.~V. {Keldysh}.
\newblock {Coulomb interaction in thin semiconductor and semimetal films}.
\newblock {\em JETP Lett.}, 29:658, 1979.

\bibitem{Cudazzo:2011a}
Pierluigi Cudazzo, Ilya~V. Tokatly, and Angel Rubio.
\newblock Dielectric screening in two-dimensional insulators: Implications for
  excitonic and impurity states in graphane.
\newblock {\em Phys. Rev. B}, 84:085406, 2011.

\bibitem{Chernikov:2014a}
Alexey Chernikov, Timothy~C. Berkelbach, Heather~M. Hill, Albert Rigosi, Yilei
  Li, Ozgur~Burak Aslan, David~R. Reichman, Mark~S. Hybertsen, and Tony~F.
  Heinz.
\newblock Exciton binding energy and nonhydrogenic $\mbox{R}$ydberg series in
  monolayer $\mbox{WS}_{2}$.
\newblock {\em Phys. Rev. Lett.}, 113:076802, 2014.

\bibitem{PhysRevB.98.125308}
Dinh Van~Tuan, Min Yang, and Hanan Dery.
\newblock Coulomb interaction in monolayer transition-metal dichalcogenides.
\newblock {\em Phys. Rev. B}, 98:125308, 2018.

\bibitem{1971JETPL..13..229G}
E.~F. {Gross}, V.~I. {Perel'}, and R.~I. {Shekhmamet'ev}.
\newblock {Inverse Hydrogenlike Series in Optical Excitation of Light Charged
  Particles in a Bismuth Iodide (BiI$_{3}$) Crystal}.
\newblock {\em JETP Lett.}, 13:229, 1971.

\bibitem{PhysRevB.98.115137}
Yasha Gindikin and Vladimir~A. Sablikov.
\newblock Spin-orbit-driven electron pairing in two dimensions.
\newblock {\em Phys. Rev. B}, 98:115137, 2018.

\bibitem{EFROS1984883}
Al.L. Efros and B.L. Gelmont.
\newblock Exciton dispersion law in diamond-like semiconductors.
\newblock {\em Solid State Communications}, 49(9):883 -- 884, 1984.

\bibitem{ll3_eng}
L.~D. Landau and E.~M. Lifshitz.
\newblock {\em Quantum Mechanics: Non-Relativistic Theory}.
\newblock Butterworth-Heinemann, Oxford, 1977.

\bibitem{PhysRevB.89.041405}
{Brian Skinner, B.~I. Shklovskii, and M.~B. Voloshin.
\newblock Bound state energy of a coulomb impurity in gapped bilayer graphene.
\newblock {\em Phys. Rev. B}, 89:041405, 2014.}

\bibitem{baldereschi73}
A.~Baldereschi and N.O. Lipari.
\newblock Spherical model of shallow acceptor states in semiconductors.
\newblock {\em Phys. Rev. B}, 8:2697, 1973.

\bibitem{PhysRevLett.115.027402}
J.~Thewes, J.~Heck\"otter, T.~Kazimierczuk, M.~A\ss{}mann, D.~Fr\"ohlich,
  M.~Bayer, M.~A. Semina, and M.~M. Glazov.
\newblock Observation of high angular momentum excitons in cuprous oxide.
\newblock {\em Phys. Rev. Lett.}, 115:027402, Jul 2015.

\bibitem{PhysRevB.98.235401}
{M.~Van~der Donck and F.~M. Peeters.
\newblock Excitonic complexes in anisotropic atomically thin two-dimensional
  materials: Black phosphorus and {TiS}$_{3}$.
\newblock {\em Phys. Rev. B}, 98:235401, 2018.}

\bibitem{PhysRevLett.96.126402}
{A.~V. Chaplik and L.~I. Magarill.
\newblock Bound states in a two-dimensional short range potential induced by
  the spin-orbit interaction.
\newblock {\em Phys. Rev. Lett.}, 96:126402, 2006.}

\bibitem{mccann:161403}
{Edward McCann.
\newblock Asymmetry gap in the electronic band structure of bilayer graphene.
\newblock {\em Phys. Rev. B}, 74(16):161403, 2006.}

\bibitem{PhysRevB.105.125404}
A.~Tiene, J.~Levinsen, J.~Keeling, M.~M. Parish, and F.~M. Marchetti.
\newblock Effect of fermion indistinguishability on optical absorption of doped
  two-dimensional semiconductors.
\newblock {\em Phys. Rev. B}, 105:125404, 2022.

\bibitem{Li:2020aa}
Zhipeng Li, Tianmeng Wang, Shengnan Miao, Zhen Lian, and Su-Fei Shi.
\newblock Fine structures of valley-polarized excitonic states in monolayer
  transitional metal dichalcogenides.
\newblock 9(7):1811--1829, 2023-01-04 2020.

\bibitem{He:2020aa}
Minhao He, Pasqual Rivera, Dinh Van~Tuan, Nathan~P. Wilson, Min Yang, Takashi
  Taniguchi, Kenji Watanabe, Jiaqiang Yan, David~G. Mandrus, Hongyi Yu, Hanan
  Dery, Wang Yao, and Xiaodong Xu.
\newblock Valley phonons and exciton complexes in a monolayer semiconductor.
\newblock {\em Nature Communications}, 11(1):618, 2020.

\bibitem{Robert:2021wc}
Cedric Robert, Sangjun Park, Fabian Cadiz, Laurent Lombez, Lei Ren, Hans
  Tornatzky, Alistair Rowe, Daniel Paget, Fausto Sirotti, Min Yang, Dinh
  Van~Tuan, Takashi Taniguchi, Bernhard Urbaszek, Kenji Watanabe, Thierry
  Amand, Hanan Dery, and Xavier Marie.
\newblock {Spin/valley pumping of resident electrons in WSe$_2$ and WS$_2$
  monolayers}.
\newblock {\em Nature Communications}, 12(1):5455, 2021.

\bibitem{Sergeev:2001aa}
R.A. Sergeev and R.A. Suris.
\newblock Ground-state energy of $x^-$ and $x^+$ trions in a two-dimensional
  quantum well at an arbitrary mass ratio.
\newblock {\em Physics of the Solid State}, 43(4):746--751, 2001.

\bibitem{suris:correlation}
R.~A. Suris.
\newblock {\em Optical Properties of 2D Systems with Interacting Electrons}, ed. by W. Ossau and R. Suris,
  chap. {Correlation between trion and hole in Fermi distribution in process
  of trion photo-excitation in doped QWs}.
\newblock NATO ASI, 2003.

\bibitem{2019arXiv191204873F}
Christian Fey, Peter Schmelcher, Atac Imamoglu, and Richard Schmidt.
\newblock Theory of exciton-electron scattering in atomically thin
  semiconductors.
\newblock {\em Phys. Rev. B}, 101:195417, 2020.

\bibitem{PhysRevB.103.075417}
Dmitry~K. Efimkin, Emma~K. Laird, Jesper Levinsen, Meera~M. Parish, and
  Allan~H. MacDonald.
\newblock Electron-exciton interactions in the exciton-polaron problem.
\newblock {\em Phys. Rev. B}, 103:075417, 2021.

\bibitem{Ju:2017aa}
{Long Ju, Lei Wang, Ting Cao, Takashi Taniguchi, Kenji Watanabe, Steven~G.
  Louie, Farhan Rana, Jiwoong Park, James Hone, Feng Wang, and Paul~L. McEuen.
\newblock Tunable excitons in bilayer graphene.
\newblock {\em Science}, 358(6365):907--910, 2017.}

\bibitem{PhysRevB.90.235302}
Dmitry~V. Rybkovskiy, Alexander~V. Osadchy, and Elena~D. Obraztsova.
\newblock {Transition from parabolic to ring-shaped valence band maximum in
  few-layer GaS, GaSe, and InSe}.
\newblock {\em Phys. Rev. B}, 90:235302,  2014.

\bibitem{PhysRevLett.96.026402}
B.~Arnaud, S.~Leb\`egue, P.~Rabiller, and M.~Alouani.
\newblock Huge excitonic effects in layered hexagonal boron nitride.
\newblock {\em Phys. Rev. Lett.}, 96:026402,  2006.

\bibitem{Cassabois:2016aa}
G.~Cassabois, P.~Valvin, and B.~Gil.
\newblock Hexagonal boron nitride is an indirect bandgap semiconductor.
\newblock {\em Nature Photonics}, 10:262,  2016.

\bibitem{Altarelli19781101}
M.~Altarelli, R.~A. Sabatini, and N.~O. Lipari.
\newblock {Camel's back excitons in GaP}.
\newblock {\em Solid State Communications}, 25(12):1101 -- 1104, 1978.

\bibitem{Glinskii1979631}
G.~F. Glinskii, A.~A. Kopylov, and A.~N. Pikhtin.
\newblock {Indirect exciton dispersion in III-V semiconductors: `Camel's back'
  in GaP}.
\newblock {\em Solid State Communications}, 30(10):631 -- 634, 1979.

\end{thebibliography}

\end{document}